\documentclass[aps,prd,preprint,onecolumn,groupedaddress]{revtex4-1}
\usepackage{amsmath}
\usepackage{amsfonts}
\usepackage{amssymb}

\begin{document}
\title{Avoiding instabilities in antisymmetric tensor field driven inflation}

\author{Sandeep Aashish}
\email[]{sandeepa16@iiserb.ac.in}

\author{Abhilash Padhy}
\email[]{abhilash92@iiserb.ac.in}

\author{Sukanta Panda}
\email[]{sukanta@iiserb.ac.in}

\affiliation{Department of Physics, Indian Institute of Science Education and Research, Bhopal 462066, India}

\date{August 29, 2019}

\begin{abstract}
Models of inflation with antisymmetric tensor studied in the past are plagued with ghost instability even in an unperturbed FRW background. We show that it is possible to avoid ghosts in an unperturbed FRW background by considering the most general kinetic term for antisymmetric tensor field. The kinetic part acquires a new gauge symmetry violating term whose effect on perturbed modes is to prevent the appearance of nondynamical modes, and thus avoid ghosts. For completeness, we perform a check for gradient instability and derive the conditions for perturbations to be free of gradient instability.
   \end{abstract}
   \maketitle
   
   \section{Introduction}
Inflation as a paradigm to explain horizon and flatness problem of early universe was first introduced by Guth \cite{guth1981}, and since has led to more than three decades of effort to build models of inflation that fit well with the observed CMB data (see Ref. \cite{martin2016} for a review). With the advent of high-precision observational data (like the recent Planck 2018 results \cite{planck2018x}), majority of scalar field driven inflation models have been ruled out while the ones in agreement are tightly constrained. More recently, new set of theoretical conditions called the Swampland criteria arise from the requirements for any effective field theory to admit string theory UV completion \cite{brennan2018,andriot2018,garg2018,obied2018,kallosh2019}, and further constrain scalar field potentials. There is thus a genuine interest to explore inflationary scenario with alternative driving fields. Some major programs include multiple fields, vector and/or gauge fields. For a comprehensive review, see Ref. \cite{martin2014}.
   
Among the theories not involving scalar fields, in particular those with vector fields \cite{ford1989,burd1991,golovnev2008,darabi2014,bertolami2015}, constructing successful models is often marred by ghost and gradient instabilities \cite{peloso2009,ryonamba2017} that lead to unstable vacua. Inflation with non-Abelian gauge fields have been shown to be free from these instabilities \cite{jabbari2011,jabbari2012,jabbari2013a,jabbari2013b}, but are in tension with Planck data and hence ruled out \cite{peloso2013}. Our endeavour is to explore inflation models with rank-2 antisymmetric tensor fields. Also referred to as the Kalb-Ramond fields, they appear naturally in the low energy limit of superstring models \cite{rohm1986,ghezelbash2009}. There are no observational signatures of antisymmetric fields in the present universe \cite{das2018}, but it is interesting to study them in the early universe when their presence may become significant\cite{elizalde2018}. 
   
Past attempts at studying inflation with antisymmetric tensor have not been successful because of the possibility of ghosts as a generic feature of the theory \cite{koivisto2009a,aashish2018c}. Even with an unperturbed Friedmann Lemaître Robertson Walker (FLRW) metric background, the perturbations to field components admit ghosts and this result remains unaffected for different choices of couplings and potential. The cause of this instability can be traced to the presence of nondynamical modes for some components of the field, which in turn is due to the structure of the gauge invariant kinetic term in these models. It turns out that the choice of kinetic term is indeed not general \cite{altschul2010}, and one can in principle consider a model with modifications to the kinetic part of action. 
   
In this work, we show that by working with a general kinetic term, it is possible to avoid ghost and gradient instabilities in antisymmetric tensor driven inflation in an unperturbed FLRW metric. The most general kinetic term for an even-parity antisymmetric tensor $B_{\mu\nu}$ upto quadratic order in field components and derivatives, is
\begin{eqnarray}
   \label{intro1}
   c_{1}\nabla_{\lambda}B^{\mu\nu}\nabla^{\lambda}B_{\mu\nu} + c_{2}\nabla_{\lambda}B^{\mu\nu}\nabla^{\mu}B_{\nu\lambda} + c_{3}\nabla_{\lambda} B^{\lambda \nu}\nabla_{\mu} B^{\mu }_{\ \nu},
   \end{eqnarray}
   which is equivalent to,
   \begin{eqnarray}
   \label{intro2}
   c_{4}H_{\lambda \mu \nu} H^{\lambda \mu \nu} + c_{5}\nabla_{\lambda} B^{\lambda \nu}\nabla_{\mu} B^{\mu }_{\ \nu}
   \end{eqnarray}
   upto some constant coefficients $c_{i}$. The first term in Eq. (\ref{intro2}) is the standard gauge invariant kinetic term, while the second term is a new non-gauge-invariant term and is taken into account in the present analysis. In fact, consideration of gauge-violating kinetic terms is not new in literature. Several vector and antisymmetric tensor field models with gauge-violating kinetic terms have been studied in the past extensively in the context of spontaneous Lorentz violation \cite{bluhm2008b,bluhm2009,hernaski2016}. 
   
   The organization of this letter is as follows. In Sec. \ref{sec2}, we study the effect of modifying the kinetic part on the background cosmology for a particular choice of background structure of $B_{\mu\nu}$. We then study perturbations to $B_{\mu\nu}$ and subsequently the ghost and gradient instability in an unperturbed FRW spacetime, in Sec. \ref{sec3}. We conclude with a few remarks on future directions in Sec. \ref{sec4}.
   
   \section{\label{sec2}Background Cosmology} 
   We begin by briefly reviewing the results of Ref. \cite{aashish2018c}, where the authors first considered the possibility of inflation driven by a rank-2 antisymmetric tensor. A typical action for antisymmetric inflation model has the form, 
   \begin{eqnarray}
   \label{bga0}
   S_{I} = \int d^{4}x\sqrt{-g}\left(- \frac{1}{12} H_{\lambda \mu \nu} H^{\lambda \mu \nu} - V(B) + \mathcal{L}_{NM} \right).
   \end{eqnarray}
   where $V(B)$ is the potential, which in our case is quadratic, $m^{2}B_{\mu\nu}B^{\mu\nu}/4$,  and $\mathcal{L}_{NM}$ is a nonminimal coupling term. the metric signature ($- + + +$). $H_{\lambda \mu \nu}(B) = \nabla_{\lambda} B_{\mu \nu} + \nabla_{\mu} B_{\nu \lambda} + \nabla_{\nu} B_{\lambda \mu }$ ($\nabla_{\mu}$ is the covariant derivative) constitutes the kinetic term and admits gauge invariance under the transformation 
   \begin{eqnarray}
   \label{bga1}
   B_{\mu\nu}\longrightarrow B^{\xi}_{\mu\nu} = B_{\mu\nu} + \nabla_{\mu}\xi_{\nu}-\nabla_{\nu}\xi_{\mu}.
   \end{eqnarray}
   
   A peculiar characteristic of antisymmetric tensor models is that while minimally coupled models generically fail to support inflation, those with nonminimal coupling can give rise to stable de-Sitter solutions and support slow-roll inflation. The choice of nonminimal coupling term does not affect the extent of support for inflation and is only restricted by theoretical constraints like stability near a Schwarzschild metric \cite{prokopec2006}. Specifically, upto quadratic order in $B_{\mu\nu}$ and second order metric derivative, allowed choices for $\mathcal{L}_{NM}$ are $B^{\mu \nu} B_{\mu \nu} R$ and $B^{\lambda \nu} B^{\mu}_{\ \nu} R_{\lambda \mu}$. However, for any choice of $\mathcal{L}_{NM}$ in action (\ref{bga0}) the perturbations to $B_{\mu\nu}$ in FLRW background admit ghosts \cite{aashish2018c,koivisto2009a} induced by the presence of nondynamical modes of perturbation. This rather generic problem has hindered the progress towards building inflation models with antisymmetric tensor, and remains to be addressed before any serious effort for analysing the full perturbation theory, including metric perturbations.  
   
   Although all possible couplings upto quadratic order have been exhausted, and it might be tempting to explore higher order couplings of $B_{\mu\nu}$ and $R$ for a resolution to ghosts, modifications to the kinetic term of action (\ref{bga0}) as yet remain unexplored. Therefore, we start with constructing the most general kinetic term upto quadratic order in $B_{\mu\nu}$, which yields a new gauge-symmetry breaking kinetic term in addition to the gauge invariant kinetic term already present in action (\ref{bga0}) \cite{altschul2010},
   \begin{eqnarray}
   \label{bga2}
   \nabla_{\lambda} B^{\lambda \nu}\nabla_{\mu} B^{\mu }_{\ \nu}.
   \end{eqnarray}
   The action that we work with is then given by,
   \begin{eqnarray}
   \label{bga3}
   S = \int d^4x \sqrt{-g} \Big[- \frac{1}{12} H_{\lambda \mu \nu} H^{\lambda \mu \nu} + \dfrac{\tau}{2} (\nabla_{\lambda} B^{\lambda \nu})(\nabla_{\mu} B^{\mu }_{\ \nu}) + (\frac{\xi}{2\kappa}R - \frac{m^2}{4}) B^{\mu \nu}B_{\mu \nu} \nonumber \\ + \frac{\zeta}{2\kappa} B^{\lambda \nu }B^{\mu}_{\ \nu} R_{\lambda \mu}  \Big].
   \end{eqnarray}
   Of course, a whole new class of terms arise if one also takes into account the parity-odd dual tensor $\mathcal{B}_{\mu\nu}$, defined by \cite{altschul2010}
   \begin{eqnarray}
   \label{bga4}
   \mathcal{B}_{\mu\nu}\equiv \dfrac{1}{2}\epsilon_{\mu\nu\rho\sigma}B^{\rho\sigma}.
   \end{eqnarray} 
   But we restrict ourselves to only parity-even terms for the sake of simplicity and because our goal is to show that it is indeed possible to avoid instabilities in models with antisymmetric tensor. 
   
   Apart from ghost instability, inflationary solutions are prone to gradient instability, which occurs when the speed of sound becomes imaginary. Gradient (in)stability has not been checked explicitly for the model(s) (\ref{bga0}) before. For completeness, the gradient instability check has been performed for action (\ref{bga3}) in later part of this work, albeit in a relevant limit suited to check the effect of $\tau$ term. 
   
   The background metric $g_{\mu\nu}$ is FLRW, with its components given by, 
    \begin{eqnarray}
    \label{bga5}
    g_{00} = -1 , \quad g_{i j} = a(t)^2 \delta_{ij}.
    \end{eqnarray}
   Our choice of the background structure of $B_{\mu \nu}$ is motivated by the spacetime symmetries as well as calculational convenience, and is given by
    \begin{eqnarray}
    \label{bga6}
    B_{\mu \nu} = 
    \begin{bmatrix}
    0 & 0 & 0 & 0\\
    0 & 0 & B(t) & -B(t)\\
    0 & -B(t) & 0 & B(t)\\
    0 & B(t) & -B(t) & 0
    \end{bmatrix},
    \end{eqnarray}
   along with a rescaling $B(t) = a(t)^{2}\phi(t)$, where $a(t)$ is the scale factor.   
   
   The contribution of $\tau$ term (second in Eq. (\ref{bga3})) to the background cosmology is through the modifications in Einstein equation \textit{viz.} the corresponding energy-momentum tensor, $T^{\tau}_{\mu\nu}$, given by
    \begin{eqnarray}
    \label{bga7}
    T^{\tau}_{\mu \nu} = -\dfrac{2}{\sqrt{-g}}\dfrac{\delta S_{\tau}}{\delta g_{\mu\nu}} = \dfrac{\tau}{2} \Big[ g_{\mu \nu} \left(  (\nabla_{\lambda} B^{ \sigma \lambda })(\nabla_{\rho} B^{\rho }_{\ \sigma}) + 2 B^{\sigma \lambda} \nabla_{\lambda} \nabla_{\rho} B^{\rho}_{\ \sigma} \right) + 2 (\nabla_{\lambda} B^{\lambda}_{\ \mu})(\nabla_{\rho} B^{\rho}_{ \ \nu})\nonumber\\
     + 2\left( B_{\mu}^{\ \lambda} \nabla_{\lambda} \nabla_{\rho} B_{\nu}^{\ \rho} +  B_{\nu}^{\ \lambda} \nabla_{\lambda} \nabla_{\rho} B_{\mu}^{\ \rho} \right)  \Big].
    \end{eqnarray}
   Remarkably, upon substituting the background value of the metric and $B_{\mu \nu}$  in Eq.(\ref{bga7}), one finds that
    \begin{eqnarray}
    \label{bacb3}
    T^{\tau}_{\mu \nu}(B) = 0.
    \end{eqnarray}
   This implies, there is no additional contribution to the background cosmology of action (\ref{bga0}) and all results for theory (\ref{bga0}) follow from Ref. \cite{aashish2018c}, leading to the following conclusions:
   \begin{itemize}
   \item[(i)] de-Sitter solutions exist, and 
   \item[(ii)] Slow roll inflation is supported.
   \end{itemize}
   As a side note, we point out that the vanishing $T^{\tau}_{\mu \nu}$ is specific to the choice of background Eq. (\ref{bga6}). It is certainly of academic interest to check for other choices of background, and we leave it as a future project.

  %%%%%%%%%%%%%%%%%%%%%%%%%%%%%%%%%%%%%%%%%%%%%%%%%%%%%%%%%%%%%%%%%
  
\section{\label{sec3}Perturbations}
The interesting part however is when $B_{\mu\nu}$ is perturbed. Surely, the perturbed modes have nontrivial contributions from the $\tau$ term, as we shall see. A full perturbation analysis, where perturbations to both metric and field are considered, is ideally required to investigate the viability of an inflation theory. However, as a starting point and because of the complexity of full perturbation theory (involving a total of $10$(metric) $+ 6$(field) $= 16$ perturbed modes), it is useful to check the stability of just the field perturbations while keeping the metric unperturbed. In several past studies, instabilities have been found at this stage \cite{aashish2018c, koivisto2009a}. 

Adding a perturbation $\delta B_{\mu\nu}$ to $ B_{\mu\nu}$, the perturbed action has the form,
\begin{eqnarray}
\label{pba0}
S[B+\delta B] &=& S[\delta B^{0}] + S[\delta B^{1}] + S[\delta B^{2}] \nonumber \\
&\equiv & S_{0} + S_{1} + S_{2},
\end{eqnarray}
where, terms have been segregated according to the order of perturbations. We are interested in the part of action that is quadratic in perturbations, $S_{2}$, since it leads to evolution equations of perturbed modes. Another trick that we use for our convenience is to Fourier transform the spatial part of all modes $\delta B_{\mu\nu}$,
\begin{eqnarray}
\label{pba1}
\delta B_{\mu\nu}(t,\vec{x})=\int \dfrac{d^{3}k}{(2\pi)^3}e^{-i\vec{k}\cdot\vec{x}}\delta \tilde{B}_{\mu\nu}(t,\vec{k}),
\end{eqnarray}
so that all spatial derivatives in the action get replaced by algebraic factors of $k$. In our calculations, we also utilize the freedom to choose the coordinate axis ($z-$axis) along momentum vector $\vec{k}$ so that all spatial derivatives along $x-$ and $y-$axes vanish. As a notation, throughout this paper, the coordinate ($\vec{x}$), time ($t$) and momenta ($\vec{k}$) dependence of all perturbed modes and their Fourier transforms are understood but not explicitly displayed, to save space. The resulting quadratic part of action, $\tilde{S}_{2}$, in general has a form,
\begin{eqnarray}
\label{pba2}
\tilde{S}_{2} = \int dt d^{3}k \sqrt{-g}\tilde{\mathcal{L}}_{2},
\end{eqnarray}
where $\tilde{\mathcal{L}}_{2}$ is the corresponding Lagrangian density expressed in terms of Fourier transformed modes $\delta \tilde{B}_{\mu\nu}$.

There are a total of six modes of perturbation to the field $B_{\mu\nu}$, which we represent as,
\begin{eqnarray}
\label{pba3}
\delta B_{\mu\nu} = \begin{bmatrix}
0 & -E_{1} & -E_{2} & -E_{3} \\
E_{1} & 0 & M_{3} & -M_{2} \\
E_{2} & -M_{3} & 0 & M_{1} \\
E_{3} & M_{2} & -M_{1} & 0
\end{bmatrix}.
\end{eqnarray}
Each of the perturbations $E_{i}$ and $M_{i}$ ($i=1,2,3$) form the componenents of two vectors $\vec{E}$ and $\vec{M}$ respectively, which are obtained after the time-space decomposition of $\delta B_{\mu\nu}$ \cite{altschul2010}. 

 With the substitution of the Eq. (\ref{pba3}) in the action (\ref{bga3}), the quadratic (in perturbation) part of the action, $S_2$, is given by,
 \begin{eqnarray}
     \label{pba4}
     S_2[\vec{E},\vec{M}] = \int d^4 x \ \Bigg[ \frac{1}{2 a} \dot{\vec{M}}\cdot\dot{\vec{M}} + 
\frac{\tau a}{2} \dot{\vec{E}} \cdot \dot{\vec{E}} +    
      \dfrac{\dot{\vec{M}}\cdot(\vec{\nabla} \times \vec{E})}{a} +
\tau \left( aH \dot{\vec{E}} \cdot \vec{E} - \dfrac{\dot{\vec{E}} \cdot (\vec{\nabla} \times \vec{M})}{a} \right) \nonumber \\            
      + \frac{\tau}{2a}\left( \dfrac{(\vec{\nabla} \times \vec{M}) \cdot(\vec{\nabla} \times \vec{M}) }{a^2} - 2 H \vec{E} \cdot (\vec{\nabla} \times \vec{M}) -(\vec{\nabla} \cdot \vec{E})^2   \right)\nonumber\\
     + \frac{1}{2a} \left((\vec{\nabla} \times \vec{E}) \cdot(\vec{\nabla} \times \vec{E})   - 
     \frac{1}{ a^2}(\vec{\nabla}\cdot\vec{M})^2\right)
     - \alpha_1 a(\vec{E}\cdot\vec{E}) 
    + \alpha_2 \dfrac{(\vec{M}\cdot\vec{M})}{a}  \Bigg],
     \end{eqnarray}
     where $\alpha_1$ and $\alpha_2$ are the short hand notations for the coefficients of the non derivative terms in the action,
      \begin{eqnarray}
     \label{pba5}
     \alpha_1 &=& \frac{(6\xi + 2\zeta)}{\kappa} \dot{H} + \frac{(12 \xi + 3 \zeta)}{\kappa} H^2 - \frac{\tau}{2} H^2 - \frac{m^2}{2}, \nonumber\\
      \alpha_2 &=& \frac{(6\xi - \zeta)}{\kappa} \dot{H} + \frac{(12 \xi - 3 \zeta)}{\kappa} H^2 - \frac{m^2}{2}.
        \end{eqnarray}
   The vectors $\vec{E}$ and $\vec{M}$ can be further decomposed into a curl free and a divergence free part in the following way:
      \begin{eqnarray}
     \label{pba6}
     \vec{E} =\vec{\nabla}u+ \vec{U} , \quad  
     \vec{M} = \vec{\nabla} v + \vec{V};
     \end{eqnarray}
     where, $\vec{U}$ and $\vec{V}$ are two divergence-free vector fields i.e ($\nabla_i U_i = \nabla_i V_i = 0$), whereas $u$ and $v$ are scalar fields. It can be shown that using Eq. (\ref{pba6}) in Eq. (\ref{pba4}), the scalar and vector parts of decomposition (\ref{pba6}) get decoupled, and $S_2$ can be written as,
      \begin{eqnarray}
     \label{pba7}
     S_2[\vec{E},\vec{M}] = S_{scalar}[u,v] + S_{vec}[\vec{U},\vec{V}],
     \end{eqnarray}
  where, 
  \begin{eqnarray}
     \label{pba8}
     S_{scalar}[u,v] = \int d^4x \Big[ -\frac{1}{2a}\left( \dot{v}\nabla^2 \dot{v} + \tau a^2 \dot{u}\nabla^2 \dot{u}   \right) - \tau a H \dot{u}\nabla^2 u \nonumber\\ 
     - \frac{ (1+ \tau a^2) }{2a^3}(\nabla^2 u)^2\nonumber\\
     + \left(\alpha_1 a u\nabla^2 u - \alpha_2 \dfrac{v\nabla^2 v}{a}  \right)  \Big];\\
     \label{pba9}
     S_{vec}[\vec{U},\vec{V}] = \int d^4x \Big[ 
\frac{1}{2a} (\dot{\vec{V}} \cdot \dot{\vec{V}} + \tau a^2 \dot{\vec{U}} \cdot \dot{\vec{U}}) + \tau a H (\dot{\vec{U}} \cdot \vec{U}) + \nonumber\\
\frac{1}{a} \left( \dot{\vec{V}} \cdot (\vec{\nabla} \times \vec{U}) - \tau \dot{\vec{U}} \cdot (\vec{\nabla} \times \vec{V}) \right)\nonumber\\
- \frac{\tau H}{a} \vec{U} \cdot (\vec{\nabla} \times \vec{V}) - \alpha_1 a \vec{U} \cdot \vec{U} + 2 \alpha_2 \frac{\vec{V} \cdot \vec{V}}{a}     
     \Big].
     \end{eqnarray}

   As described before, for the present analysis we Fourier transform $S_2$ in a suitable frame so that the momentum vector ($\vec{k}$) lies along z-axis, to obtain $\tilde{S}_2$. For convenience, we do not adopt different notations for Fourier transforms of functions since here onwards we only work in Fourier space. It turns out that in $\tilde{S}_2$, the vector part $\tilde{S}_{vec}[\vec{U},\vec{V}]$ can once again be written as a sum of two terms, $\tilde{S}_{vec}^{(1)}[U_x,V_y]$ and $\tilde{S}_{vec}^{(2)}[U_y,V_x]$, so that
   \begin{equation}
      \label{new0}
      \tilde{S}_2[\vec{E},\vec{M}] = \tilde{S}_{scalar}[u,v] + \tilde{S}_{vec}^{(1)}[U_x,V_y] + \tilde{S}_{vec}^{(2)}[U_y,V_x],
   \end{equation}
where,
   \begin{eqnarray}
\label{pba10}
\tilde{S}_{scalar}[u,v] = \int dt d^3k \  k^2 \Big[ \frac{1}{2a} (\dot{v}^{\dagger} \dot{v} + \tau a^2 \dot{u}^{\dagger} \dot{u}) + \frac{\tau a H}{2} (\dot{u}^{\dagger} u + h.c) - (\tau \frac{k^2}{2a} + a\alpha_1) u^{\dagger} u \nonumber\\
+ \frac{1}{2a} (2 \alpha_2 - k^2) v^{\dagger}v \Big] \\
\label{pba11}
\tilde{S}_{vec}^{(1)}[U_x,V_y] = \int dt d^3k \ \Big[ \frac{1}{2a} (\dot{V}_x^{\dagger} \dot{V}_x + \tau a^2 \dot{U}_y^{\dagger} \dot{U}_y ) + \frac{\tau a H}{2} (\dot{U}_y^{\dagger} U_y + h.c) \nonumber\\
+ \frac{ik}{2a} \left( (\dot{V}_x^{\dagger}U_y - h.c) + \tau (\dot{U}_y^{\dagger}V_x - h.c) - \tau H (V_x^{\dagger} U_y - h.c) \right)\nonumber\\
- a \alpha_1 U_y^{\dagger} U_y + \frac{\alpha_2}{a} V_x^{\dagger} V_x 
 \Big]
\\
\label{pba12}
\tilde{S}_{vec}^{(2)}[U_y,V_x] = \int dt d^3k \ \Big[ \frac{1}{2a} (\dot{V}_y^{\dagger} \dot{V}_y + \tau a^2 \dot{U}_x^{\dagger} \dot{U}_x ) + \frac{\tau a H}{2} (\dot{U}_x^{\dagger} U_x + h.c) \nonumber\\
- \frac{ik}{2a} \left( (\dot{V}_y^{\dagger}U_x - h.c) + \tau (\dot{U}_x^{\dagger}V_y - h.c) - \tau H (V_y^{\dagger} U_x - h.c) \right)
- a \alpha_1 U_x^{\dagger} U_x + \frac{\alpha_2}{a} V_y^{\dagger} V_y 
 \Big]
\end{eqnarray}

Our objective now is to check for the ghost and gradient instability in $\tilde{S}_2[\vec{E},\vec{M}]$. Ghosts appear in a theory whenever the kinetic term(s) acquire a negative sign, implying a negative and thus unbounded kinetic energy.  Gradient instability appears due to wrong sign before the momentum square term in the action as it leads to an unbounded Hamiltonian, and at high energies the gradient instability can act as ghost \cite{wolf2019}. In what follows, we derive conditions avoiding ghosts and gradient instability in the present model.
   
\subsection{Ghost Instability}
The relevant term for analysing ghosts in Eq. (\ref{new0}) is its kinetic part, which can be cast into the form,
\begin{eqnarray}
   \label{gi0}
   \tilde{S}^{kin}_{2} = \int dt d^{3}k \dot{\Delta}^{\dagger}T\dot{\Delta},
\end{eqnarray}
where, $\Delta$ is an array consisting of all perturbed modes, and is given by, 
\begin{eqnarray}
   \label{gi1}
   \Delta = \begin{bmatrix}
   v \\
   V_x \\
   V_y \\
   u \\
   U_y \\
   U_x
   \end{bmatrix}; \quad 
   \Delta^{\dagger} = \begin{bmatrix}
   v^{\dagger} & V_{x}^{\dagger} & V_{y}^{\dagger} & u^{\dagger} & U_{y}^{\dagger} & U_{x}^{\dagger}
   \end{bmatrix}.
\end{eqnarray}
The coefficients of kinetic terms in Eq. (\ref{new0}) are encompassed in the matrix $T$ which reads,
\begin{eqnarray}
   \label{gi2}
   T = \begin{bmatrix}
   \dfrac{k^{2}}{2a(t)} & 0 & 0 & 0 & 0 & 0\\
   0 & \dfrac{1}{2a(t)} & 0 & 0 & 0 & 0 \\
   0 & 0 &\dfrac{1}{2a(t)} & 0 & 0 & 0\\
   0 & 0 & 0 & \dfrac{k^{2}a(t)\tau}{2} & 0 & 0\\
   0 & 0 & 0 & 0 & \dfrac{a(t)\tau}{2} & 0\\
   0 & 0 & 0 & 0 & 0 & \dfrac{a(t)\tau}{2} 
   \end{bmatrix}.
   \end{eqnarray}
   Note that there are no off-diagonal terms present and there are no non-dynamical modes in $\tilde{S}_{2}$, which leads us to conclude that there are no ghosts provided that the coupling $\tau$ satisfies a simple no-ghost condition:
   \begin{eqnarray}
   \label{gi3}
   \tau > 0.
   \end{eqnarray}
   Clearly, when $\tau = 0$, modes $u, \ \vec{U}$, become nondynamical and would lead to ghosts \citep{aashish2018c}. 
 
 \subsection{Gradient Instability}
Gradient instability can be checked by evaluating the speed of sound, $c_s$ in a given theory. An imaginary value for the sound speed implies gradient instability. To calculate $c_s$ in the present model, one needs to first derive the equations of motion from action (\ref{new0}). For this purpose, we introduce three ($2\times 1$) matrices $\Delta_{i}$ ($i=1,2,3$) defined as, 
 \begin{eqnarray}
   \label{pbb2}
   \Delta_1 = 
   \begin{pmatrix}
   u\\v
   \end{pmatrix} ; \quad
   \Delta_2 = 
   \begin{pmatrix}
   U_x\\V_y
   \end{pmatrix} ; \quad
   \Delta_3 = 
   \begin{pmatrix}
   U_y\\V_x
   \end{pmatrix};
   \end{eqnarray}
   and vary Eq. (\ref{new0}) with respect to $\Delta^{\dagger}_{i}$ ($i=1,2,3$) to obtain,
 \begin{eqnarray}
\label{pbb1}
\ddot{\Delta}_1 + \Xi_1 \dot{\Delta}_1 + \Sigma_1 \Delta_1 = 0; \nonumber\\
\ddot{\Delta}_2 + \Xi_2 \dot{\Delta}_2 + \Sigma_2 \Delta_2 = 0; \\
\ddot{\Delta}_3 + \Xi_3 \dot{\Delta}_3 + \Sigma_3 \Delta_3 = 0; \nonumber
\end{eqnarray} 
and $\Xi_{i}$ and $\Sigma_{i}$ are the coefficient matrices of order $(2 \times 2)$ given by,
 \begin{eqnarray}
 \label{pbb3}
 \Xi_1 = 
 \begin{pmatrix}
 H & 0\\
 0 & -H
 \end{pmatrix} \quad \quad
 &\Sigma_1 = 
 \begin{pmatrix}
 \dot{H} + H^2 + \frac{k^2}{a^2} + \frac{2 \alpha_1}{\tau} & 0\\
 0 & k^2 - 2\alpha_2
 \end{pmatrix}\nonumber\\
 \Xi_2 = 
 \begin{pmatrix}
 H & -\frac{ik(\tau + 1)}{a^2 \tau}\\
 -ik(\tau + 1) & -H
 \end{pmatrix} \quad \quad
 & \Sigma_2 = 
 \begin{pmatrix}
 \dot{H} + H^2 + \frac{2\alpha_1}{\tau} & \frac{2ikH}{a^2}\\
 -ikH(\tau - 1) & -2\alpha_2
 \end{pmatrix}\nonumber\\
 \Xi_3 = 
 \begin{pmatrix}
 H & \frac{ik(\tau + 1)}{a^2 \tau}\\
 ik(\tau + 1) & -H
 \end{pmatrix} \quad \quad
 &\Sigma_3 = 
 \begin{pmatrix}
 \dot{H} + H^2 + \frac{2\alpha_1}{\tau} & -\frac{2ikH}{a^2}\\
 ikH(\tau - 1) & -2\alpha_2
 \end{pmatrix}
 \end{eqnarray}
 A reasonable assumption for solutions to Eq. (\ref{pbb1}) in the deep subhorizon is to take $\Delta_{i} \propto \exp[-i\int^{t}c_{si}k/a(t')dt']\vec{e}_{i}$ as the solution to eigenvector equation (\ref{pbb1}), where $c_{si}$ is the sound speed and is treated as constant ($\dot{c}_{si}\ll k$) \cite{emami2017}, and $\vec{e}_i$ is a constant vector. Substituting this ansatz in Eq.(\ref{pbb1}) and neglecting $\dot{c}_{si}$ terms, we get quadratic equations in terms of $c_{si}^2$,
  \begin{eqnarray}
 \label{pbb5}
 \Big[ c_{s1}^2 - 1- \frac{a^2}{k^2}\left( \dot{H} + H^2 + \frac{2\alpha_1}{\tau}  \right) \Big] \Big[ c_{s1}^2  - 2ic_{s1}\frac{aH}{k} -a^2 + \frac{2\alpha_2 a^2}{k^2}\Big] &=& 0; \\
 \label{pbb6}
 c_{si}^4 - 2i c_{si}^3 \frac{aH}{k} + \Big[ \frac{a^2}{k^2} \left( 2\alpha_2 - \frac{2\alpha_1}{\tau} - \dot{H} - H^2 \right) - \frac{(\tau +1)^2}{\tau} \Big]c_{si}^2 \nonumber\\
 + 2ic_{si}\frac{aH}{k} \Big[ \frac{a^2}{k^2} \left( \dot{H} + H^2 + \frac{\alpha_1}{\tau} \right) + \frac{(\tau + 1)^2}{2\tau}  \Big] - 2(\tau - 1) \frac{a^2 H^2}{k^2} &=& 0, \ i=2,3.
 \end{eqnarray}
 A theory suffers from gradient instability when the speed of sound, $c_{si}$ (defined in the relativistic fluid approximation, see Ref. \cite{dodelson2003}), becomes imaginary. Hence, to avoid gradient instability one must demand that $c_{si}^{2}>0$. Solving Eqs. (\ref{pba5}) and (\ref{pba6}) for $c_{si}^2$ will lead to conditions for avoiding gradient instability. However, our current interest is limited to deep subhorizon, where $k >> aH$. In other words, we restrict ourselves to high momentum limit. In fact, while instabilities can arise in the low momentum limit, they have been shown to be Jeans-like instabilities and may thus be under control \cite{sotiriou2016}. Solving above equations in the limit $k >> aH$ leads to, 
  \begin{eqnarray}
 \label{pbb7}
 c_{s1}^2 = 1-\frac{a^2 m^2}{k^2 \tau} \ , \
a^2(1+\frac{m^2}{k^2});
 \end{eqnarray}
 and for $j=2,3$: 
 \begin{eqnarray}
 \label{pbb9}
 c_{sj}^2 = 0, \ \frac{(\tau +1)^2}{\tau} + \frac{a^2 m^2}{k^2} \frac{(\tau - 1)}{\tau}. 
 \end{eqnarray}
 From Eqs. (\ref{pbb7}) and (\ref{pbb9}), and taking into account the positivity of $\tau$ given by Eq. (\ref{gi3}), the conditions on $\tau$ required for a positive $c_{si}^{2}$ (where $i=1,2,3$) are,
 \begin{eqnarray}
 \label{pbb8}
 \tau > \frac{a^2 m^2}{k^2},\nonumber \\
 \tau > - \left( 1+ \frac{a^2 m^2}{2k^2}\right) + \sqrt{\frac{2 a^2 m^2}{k^2}\left(1+\frac{a^2 m^2}{8k^2}\right)}.
 \end{eqnarray}
Conditions (\ref{pbb8}) along with (\ref{gi3}) constrain the parameter $\tau$ for which the theory (\ref{bga3}) is free of ghost and gradient instabilities. In fact, it is straightforward to check that in the limit $k\to\infty$, these conditions reduce to $\tau > 0$ and are trivially satisfied.

%%%%%%%%%%%%%%%%%%%%%%%%%%%%%%%%%%%%%%%%%%%%%%%%%%%%%%%%%%%%%%%%%%%%%%%%%%%%%%%%%%%%%

\section{\label{sec4}Conclusion}
We showed that by including a new kinetic ($\tau$) term in the action (\ref{bga0}) it is possible to avoid ghost instabilities in perturbations. This can be attributed to the absence of nondynamical modes that otherwise lead to ghosts \cite{koivisto2009a,aashish2018c}. For our choice of background $B_{\mu\nu}$, the $\tau$ term does not affect background cosmology. We performed gradient instability analysis for perturbations in $B_{\mu\nu}$ and derived conditions on $\tau$ to avoid gradient instability. In the high momentum limit ($k\to\infty$), the theory is trivially free from ghost and gradient instabilities for all positive $\tau$. 

The results of this analysis present a strong case for more detailed investigations of ghost, gradient and other instabilities for perturbations including the metric part, and should motivate further directions in inflation model building. The choice of kinetic term (\ref{intro2}) also motivates further analysis of the physical degrees of freedom, that can be addressed through Hamiltonian analysis using $3+1$ ADM decomposition. An important aspect of academic interest is to study the effect of different choices of background structure of $B_{\mu\nu}$. Another interesting problem is to explore the cosmology and viability of parity-odd terms, which the authors plan to pursue in future. Additionally, studies involving the higher order terms of $B_{\mu\nu}$ and gravity may also be explored.

\begin{acknowledgments}
Some manipulations in this work were done using Maple\texttrademark \footnote{Maple 2017.3, Maplesoft, a division of Waterloo Maple Inc., Waterloo, Ontario.} and most of them have been cross-checked by hand. This work is partially supported by DST (Govt. of India) Grant No. SERB/PHY/2017041. S.A. wishes to thank ICTS (Bengaluru), where a part of this work was completed, for hospitality.
\end{acknowledgments}

\bibliography{ref,references}
\end{document}